\documentclass[12pt,preprint]{aastex}

\shorttitle{Electron-positron pair production in the electrosphere
} \shortauthors{Cheng\& Harko}
\usepackage{subfigure}
\begin{document}
\title{Electron-positron pair production in the electrosphere of quark stars}
\author{T.~Harko$^1$ and K.~S.~Cheng$^2$}
\affil{Department of Physics, The University of Hong Kong, Pok Fu
Lam Road, Hong Kong, Hong Kong SAR, P. R. China}
\email{$^1$harko@hkucc.hku.hk, $^2$hrspksc@hkucc.hku.hk}

\begin{abstract}
We study Schwinger pair creation of charged particles due to the
inhomogeneous electric field created by the thin electron layer at
the surface of quark stars (the electrosphere). As suggested
earlier, due to the low photon emissivity of the quark-gluon
plasma and of the electrosphere, electron-positron pair emission
could be the main observational signature of quark stars. To
obtain the electron-positron pair creation rate we use the
tunnelling approach. Explicit expressions for the fermion creation
rate per unit time per unit volume are derived, which generalize
the classical Schwinger result. The finite size effects in pair
production, due to the presence of a boundary (the surface of the
quark star), are also considered in the framework of a simple
approach. It is shown that the boundary effects induce large
quantitative and qualitative deviations of the particle production
rate from what one deduces with the Schwinger formula and its
generalization for the electric field of the electrosphere. The
electron-positron pair emissivity and flux of the electrosphere of
quark stars due to pair creation is considered, and the magnitude
of the boundary effects for this parameters is estimated. Due to
the inhomogeneity of the electric field distribution in the
electrosphere and of the presence of the boundary effects, at high
temperatures ($T\geq T_{cr}\approx 0.1 $ MeV) we find a lower
electron-positron flux as previously estimated. The numerical
value of the critical temperature $T_{cr}$ depends on the surface
potential of the star. We briefly consider the effect of the
magnetic field on the pair creation process and show that the
magnetic field can enhance drastically the pair creation rate.
\end{abstract}


\section{Introduction}

The Schwinger mechanism of pair production \citep{Sch51}, first
proposed to study the production of electron-positron pairs in a
strong and uniform electric field, has been applied to many
problems in contemporary physics. Strong electromagnetic fields
lead to two physically important phenomena: pair production and
vacuum polarization. A strong electric field makes the quantum
electrodynamic vacuum (QED) unstable, and consequently it decays
by emitting a significant number of boson or fermion pairs
\citep{Gr85}.

For a spin $1/2$ particle Schwinger's predicted production rate
per unit time and volume $w$ is given by \citep{Sch51,So82}
\begin{equation}\label{prod}
w(E_0)=eE_0\int \frac{d^{2}k_{i}}{\left( 2\pi \right) ^{2}}\sum_{n=1}^{\infty }%
\frac{1}{n}\exp \left[ -\frac{\pi n\left( m_{e}^{2}+k_{i}^{2}\right) }{%
\left| eE_0\right| }\right] ,
\end{equation}
where $m_{e}$ and $e$ are the electron mass and charge,
respectively, $k_{i}$ is the transverse momentum and $E_0$ is the
(constant) electric field. In the present paper we use units so
that $\hbar =c =k_B =1$. In these units, $e$ is equal to
$\alpha^{1/2}$ and $1$ MeV$=5.064\times 10^{-3}$ fm$^{-1}$.


In most of the physical applications the proper time method
introduced by \citet{Sch51} has been used to calculate the pair
production rate. The real part of the effective action leads to
vacuum polarization and the imaginary part to pair production.
Though that method is conceptually well defined and technically
rigorous, it is generally difficult to apply it to concrete
physical problems such as inhomogeneous electromagnetic fields.
Schwinger's result was generalized to
electric fields $E_3=E_3\left(x_{\pm }\right)$, which depend upon
either light cone coordinates $x_{\pm }=x_3\pm x_0$,  but not upon
both, in \citet{To00}. The form of the result is exactly the same
as in the original formula given by Eq. (\ref{prod}). The case of
electric fields depending on both $x_+$ and $x_{-}$,
$E_3=E_3\left(x_+,x_{-}\right)$ was considered in \citet{Av03}.

An alternative approach to pair creation was initiated by
\citet{Ca79,Ca80}, who re-derived Schwinger's pair production rate
by semi-classical tunnelling calculations. The boson and fermion
pair production rate by strong static uniform or inhomogeneous
electric fields was derived, in terms of instanton tunnelling
through potential barriers in the space-dependent gauge, in
\citet{Kim02}.

The Schwinger mechanism for particle production in a strong and
uniform electric field for an infinite system was generalized to
the case were the strong field is confined between two plates
separated by a finite distance by \citet{Wa88}. The production
rates, obtained by solving the Klein-Gordon and Dirac equations in
a linear vector potential, can be expressed in an exact analytical
form. The numerical evaluations of the production rates have shown
large deviations from the Schwinger formula, thus indicating a
large finite size effect in particle production. The explicit
finite size corrections to the Schwinger formula for the rate of
pair production for uniform electric fields confined to a bounded
region were calculated, by using the Balian-Bloch expansion of the
Green functions, by \citet{Ma88,Ma89}.

One important astrophysical situation in which electron-positron
pair creation could play an extremely important role is the case
of the electrosphere of the quark stars
\citep{Us98a,Us98b,Us01,PaUs02}. Quark stars could be formed as a
result of a hadron-quark phase transition at high densities and/or
temperatures \citep{It70,Bo71,Wi84}. If the hypothesis of the
quark matter is true, then some neutron stars could actually be
strange stars, built entirely of strange matter \citep{Al86,Ha86}.
For a review of strange star properties, see \cite{Ch98}.

Several mechanisms have been proposed for the formation of quark
stars. Quark stars are expected to form during the collapse of the
core of a massive star after the supernova explosion, as a result
of a first or second order phase transition, resulting in
deconfined quark matter \citep{Da}. The proto-neutron star core,
or the neutron star core, is a favorable environment for the
conversion of ordinary matter into strange quark matter
\citep{ChDa}. Another possibility is that some neutron stars in
low-mass X-ray binaries can accrete sufficient mass to undergo a
phase transition to become strange stars \citep{Ch96}. This
mechanism has also been proposed as a source of radiation emission
for cosmological $\gamma $-ray bursts \citep{Ch96}. Some basic
properties of strange stars like mass, radius, collapse and
nucleation of quark matter in neutron stars cores have been also
studied \citep{Gl95,Gl97,ChHa,HaCh00,HaCh02,Ha04}.

The structure of a realistic strange star is very complicated, but
its basic properties can be described as follows \citep{Al86}.
Beta-equilibrated strange quark - star matter consists of an
approximately equal mixture of up $u$, down $d$ and strange $s$
quarks, with a slight deficit of the latter. The Fermi gas of $3A$
quarks constitutes a single color-singlet baryon with baryon
number $A$. This structure of the quarks leads to a net positive
charge inside the star. Since stars in their lowest energy state
are supposed to be charge neutral, electrons must balance the net
positive quark charge in strange matter stars. The electrons,
being bounded to the quark matter by the electromagnetic
interaction and not by the strong force, are able to move freely
across the quark surface, but clearly cannot move to infinity
because of the electrostatic attraction of quarks. For hot stars
the electron distribution could extend up to $\sim 10^{3}$ fm
above the quark surface \citep{Ke95,ChHa03,Us05}. The electron
distribution at the surface of the quark star is called
electrosphere. The effect of the color-flavor locked phase of
strange matter on the electric field in the electrosphere was
discussed by \citet{Us04}.

Photon emissivity is the basic parameter for determining
macroscopic properties of stellar type objects. \citet{Al86} have
shown that, because of very high plasma frequency near the strange
matter edge, photon emissivity of strange matter is very low. For
temperatures $T<<E_{p} $, where $E_{p}\approx 23$ MeV is the
characteristic transverse plasmon cutoff energy, the equilibrium
photon emissivity of strange matter is negligibly small, as
compared to the blackbody one. The spectrum of equilibrium photons
is very hard, with $\omega
>20$ MeV \citep{Chm91}.

The bremsstrahlung emissivity of the quark matter and of the
electrosphere have been estimated recently in
\citet{ChHa03,Ja04,HaCh05}. By taking into account the effect of
the interference of amplitudes of nearby interactions in a dense
media (the Landau-Pomeranchuk-Migdal effect) and the absorption of
the radiation in the external electron layer, the emissivity of
the quark matter could be six orders of magnitude lower than the
equilibrium blackbody radiation \citep{ChHa03}. The presence of
the electric field strongly influences the radiation spectrum
emitted by the electrosphere and strongly modifies the radiation
suppression pattern \citep{HaCh05}. All the radiation properties
of the electrons in the electrosphere essentially depend on the
value of the electric potential at the quark star surface.

The Coulomb barrier at the quark surface of a hot strange star may
also be a powerful source of $e^{+}e^{-}$ pairs, which are created
in the extremely strong electric field of the barrier. In the case
of the energy loss via the production of $e^{-}e^{+}$ pairs, the
energy flux is given by $F_{\pm }=\varepsilon _{\pm }\dot{n}$,
with $\varepsilon _{\pm }=m_{e}+T$ and
\begin{equation}\label{usov}
\dot{n }=\left( \frac{9T}{2\pi \varepsilon _{F}^{2}}\right)
\sqrt{\frac{\alpha }{\pi }}\exp \left( -2m_{e}/T\right)
n_{e}T^{2}J\left( \xi \right) \Delta r_E,
\end{equation}
where $\varepsilon _{F}=\left( \pi ^{2}n_{e}\right) ^{1/3}$ is the
Fermi energy of the electrons, $\xi =2\sqrt{\alpha /\pi }\left(
\varepsilon _{F}/T\right) $,
\begin{equation}\label{j}
J\left( \xi \right) =\frac{1}{3}\frac{ \xi ^{3}\ln \left( 1+2\xi
^{-1}\right)} {\left( 1+0.074\xi \right) ^{3}}+\frac{ \pi
^{5}}{6}\frac{ \xi ^{4}}{\left( 13.9+\xi \right) ^{4}},
\end{equation}
and $\Delta r_E$ is the thickness of the region with the strong
electric field \citep{Us98a,Us01}.

At surface temperatures of around $%
10^{11}$ K, the luminosity of the outflowing plasma may be of the order $%
\sim 10^{51}$ ergs$^{-1}$ \citep{Us98a,Us98b}. Moreover, as shown
by \citet{PaUs02}, for about one day for normal quark matter and
for up to a hundred years for superconducting quark matter, the
thermal luminosity from the star surface, due to both photon
emission and $e^{+}e^{-}$ pair production may be orders of
magnitude higher than the Eddington limit. Hence,
electron-positron pair creation due to the electric field of the
electrosphere could be one of the main observational signatures of
quark stars.

It is the purpose of the present paper to consider the Schwinger
process of pair creation in the electrosphere of the quark stars,
by systematically taking into account the main physical
characteristics of the environment in which pair production takes
place. The electric field in the electrosphere is not a constant,
as assumed in the Schwinger model, but it is a rapidly decreasing
function of the distance $z$ from the quark star surface.
Therefore, in order to realistically describe the pair production
process one must consider electron-positron creation in an
inhomogeneous electric field. To study the pair production process
we adopt the tunnelling approach and the fermion production rate
in the electric field of the electrosphere is derived. The
production rate is a local quantity, it depends on the distance to
the quark star's surface and it is a rapidly decreasing function
of $z$. The emissivity and energy flux due to pair creation at the
quark star surface is also considered. An important factor which
can reduce significantly the pair production rate is the presence
of a boundary (the quark star surface). For electric fields
perpendicular to the boundary there is a significant reduction in
the magnitude of the pair production rate, which is also
associated with an important qualitative change in the process,
which becomes a periodic function of the distance to the quark
star surface.

The present paper is organized as follows. In Section 2 we review
the basic properties of the electrosphere of quark stars. The
electron-positron rate production in the electric field of the
electrosphere is obtained in Section 3. The boundary effects are
considered in Section 4. In Section 5 we calculate the
electron-positron pair flux of the electrosphere and compare our
results with those obtained by \citet{Us98a,Us98b,Us01}. A brief
summary of our results is given in Section 6.

\section{Structure of the electrosphere of strange stars}

In the electrosphere, electrons are held to the strange quark
matter (SQM) surface by an extremely strong electric field. The
thickness of the electrosphere is much smaller than the stellar
radius $R\simeq 10^6$ cm, and a plane-parallel approximation may
be used to study its structure \citep{Us05}. In this approximation
all values depend only on the coordinate $z$, where the axis $z$
is perpendicular to the SQM surface ($z=0$) and directed outward.
To find the distributions of electrons and electric fields in the
vicinity of the SQM surface, we use a simple Thomas-Fermi model
considered by \citet{Al86} and take into account the finite
temperature effects as discussed by \citet{Ke95}.

This model can be solved exactly, and all physical quantities of
interest (chemical potential, electric field etc.) can be
expressed in an exact analytical form \citep{ChHa03,Us05}. The
chemical equilibrium in the electrosphere implies that the
electron chemical potential $\mu _e$ satisfies the condition
$-V\left(\infty \right)+\mu _e\left(\infty \right)=-V+\mu _{e}$,
where $V$ is the electrostatic potential per unit charge and
$V\left(\infty \right)$ and $ \mu _e\left(\infty \right)$ are the
values of the electrostatic potential and of the electron's
chemical potential at infinity, respectively. Since far outside
the star both $V\left(\infty \right)$ and $ \mu _e\left(\infty
\right)$ tend to zero, it follows that $\mu _{e}=V$ \citep{Al86}.

The structure of the electrosphere and the corresponding radiation
processes essentially depends on the value of $V_q$, the electric
charge density inside the quark star. When the temperature of the
quark star core drops below $10^{9}$ K, the strange matter becomes
superfluid. At this temperature quarks can form colored Cooper
pairs near the Fermi surface and become superconducting. From the
BCS theory it follows that the critical temperature $T_{c}$ at
which the transition to the superconducting state takes place is
$T_{c}=\Delta /1.76$, where $\Delta $ is the pairing gap energy
\citep{Bl}. An early estimation of $\Delta $ gave $\Delta \sim
0.1-1$ MeV \citep{early}, but some recent studies considering
instanton-induced interactions between quarks estimated $\Delta
\sim 100$ MeV \citep{all1}.

Strange quark matter in the color-flavor locked (CFL) phase of
QCD, which occurs for  $\Delta \sim 100$ MeV, could be rigorously
electrically neutral, despite the unequal quark masses, even in
the presence of the electron chemical potential \citep{all1}.
Hence, for the CFL state of quark matter $V_q=0$ and no electrons
are present inside or outside the quark star \citep{LH03}.

However, \citet{PaUs02} pointed out that for sufficiently large
$m_{s}$ the low density regime is rather expected to be in the
''2-color-flavor Superconductor'' phase in which only $u$ and $d$
quarks of two color are paired in single condensate while the ones
of the third color, and $s$ quarks of all three colors, are
unpaired. In this phase, some electrons are still present. In
other words, electrons may be absent in the core of strange stars
but present, at least, near the surface where the density is
lowest. Nevertheless, the presence of the CFL effect can reduce
the electron density at the surface and hence it can also
significantly reduce the intensity of the electric field and the
electromagnetic emissivity of the electrons in the electrosphere.
Therefore, in order to describe the radiation properties of the
electrosphere we assume that $V_q\neq 0$.

The Poisson equation for the electrostatic potential $V\left(
z,T\right) $ generated by the finite temperature electron
distribution reads \citep{Al86,Ke95}
\begin{equation}
\frac{d^{2}V}{dz^{2}}=\frac{4\alpha }{3\pi }\left[ \left(
V^{3}-V_{q}^{3}\right) +\pi ^{2}\left( V-V_{q}\right) T^{2}\right]
,z\leq 0, \label{p2}
\end{equation}
\begin{equation}
\frac{d^{2}V}{dz^{2}}=\frac{4\alpha }{3\pi }\left( V^{3}+\pi
^{2}T^{2}V\right) ,z\geq 0,  \label{3}
\end{equation}
where $T$ is the temperature of the electron layer, which can be
taken as a constant, since we assume the electrons are in
thermodynamic equilibrium with the constant temperature quark
matter. In Eqs.~(\ref{p2})-(\ref{3}), $z$ is the space coordinate
measuring height
above the quark surface, $\alpha $ is the fine structure constant and $%
V_{q}/3\pi ^{2}$ is the quark charge density inside the quark
matter. The boundary conditions for Eqs.~(\ref{p2})-(\ref{3}) are
$V\rightarrow V_{q}$ as $z\rightarrow -\infty $ and $V\rightarrow
0$ for $z\rightarrow \infty $. In the case of the zero temperature
electron distribution at the boundary $z=0$ we have the condition
$V(0)=(3/4)V_{q}$ \citep{Al86}.

The general solution of Eq.~(\ref{3}) is given by \citep{ChHa03}
\begin{equation}
V\left( z,T\right) =\frac{\sqrt{2}\pi T}{\sinh \left[
2\sqrt{\frac{\alpha \pi }{3}}T\left( z+z_{0}\right) \right] },
\end{equation}
where $z_{0}$ is a constant of integration. Its value can be
obtained from the condition of the continuity of the potential
across the star's surface, requiring $V_{q}(0,T)=V\left(
0,T\right) $, where $V_{q}\left( z,T\right) $ is the value of the
electrostatic potential in the region $z\leq 0$, described by
Eq.~(\ref{p2}). Therefore
\begin{equation}
z_{0}=\frac{1}{2}\sqrt{\frac{3}{\alpha \pi }}\frac{1}{T} {\rm
arcsinh}\left[ \frac{\sqrt{2}\pi T}{V_{q}\left( 0,T\right)
}\right] ,
\end{equation}

The number density distribution $n_{e}$ of the electrons at the
quark star surface can be obtained from $n_{e}\left( z,T\right)
=V^{3}/3\pi ^{2}+VT^{2}/3$ \citep{Ke95,ChHa03} and is given by
\begin{equation}
n_{e}\left( z,T\right) =\frac{\sqrt{2}\pi }{3}\frac{1+\cosh ^{2}\left[ 2%
\sqrt{\frac{\alpha \pi }{3}}T\left( z+z_{0}\right) \right] }{\sinh
^{3}\left[ 2\sqrt{\frac{\alpha \pi }{3}}T\left( z+z_{0}\right)
\right] }T^{3}.
\end{equation}

In the limit of zero temperature, $T\rightarrow 0$ we obtain
$V(z)=a_0/(z+b)$, where $a_0=\sqrt{3\pi /2\alpha }$ and $b$ is an
integration constant. $b$ can be determined from the boundary
condition $V(0)=(3/4)V_{q}$, which gives $b=\left(
4a_0/3V_{q}\right) $. Therefore, in this case we find for the
electron particle number distribution the expression:
$n_{e}(z)=(1/3\pi ^{2})a^{3}_{0}/ \left(z+b\right) ^{3}$.

In the absence of a crust of the quark star, the electron layer
can extend to several thousands fermis outside the star's surface.

The strength of the electric field $E$ outside the quark star
surface is given by
\begin{equation}
E\left( z,T\right) =\sqrt{\frac{4}{3\pi
}}V\sqrt{\frac{V^{2}}{2}+\pi ^{2}T^{2}},
\end{equation}
and can be expressed as
\begin{equation}\label{el}
E\left( z,T\right) =\sqrt{\frac{8\pi ^{3}}{3}}\frac{\cosh \left[ 2%
\sqrt{\frac{\alpha \pi }{3}}T\left( z+z_{0}\right) \right] }{\sinh
^{2}\left[ 2\sqrt{\frac{\alpha \pi }{3}}T\left( z+z_{0}\right)
\right] }T^{2}.
\end{equation}

The ratio of the electric field of the electrosphere of quark
stars and of the critical electric field $E_{cr}=m_e^2/e$ is
represented, as a function of the distance from the quark star
surface and for different values of the temperature, in Fig. 1.

\section{Electron-positron pair production rates in the electrosphere}

Let us consider a quantized Dirac field coupled to a classical
external electromagnetic field described by the potential $A_{\mu
}$. The probability $P$ to remain in the ground state, i.e., the
probability of emitting no pairs, is given by $P=\left|
\left\langle 0\left| S\right| 0\right\rangle
\right| ^{2}$, where $S$ is the $S$ matrix defined as $S=\hat{T}\exp \left[ i\int L_{I}d^{4}x%
\right] $, $\hat{T}$ is the time ordering operator and $L_{I}$ is
the interaction Lagrangian \citep{So82}. The probability $P$ can also be written as $P=\exp %
\left[ -\int w(x)d^{4}x\right] $, where $w\left( x\right) =2{\rm Im}%
L_{eff}\left( x\right) $, where $L_{eff}\left( x\right) $ is the
one-loop effective Lagrangian density, which includes all orders
in the external
field, but neglects self-interactions of the matter fields. The quantity $%
w(x)$ can be interpreted as the pair production rate per unit time
and unit volume at the space-time point $x=\left(
x_{0},x_{1},x_{2},x_{3}\right) $ \citep{Gr85,Ma89}.

An alternative description of the pair creation process can be
obtained by assuming that the vacuum decays if there exists an
ingoing antiparticle mode which is at the same time an outgoing
particle mode. An in-vacuum is defined via the ingoing
(anti)-particle basis, $_{-}a_{l}\left| {\rm in},{\rm
vac}\right\rangle =_{+}a_{k}\left| {\rm in},{\rm vac}\right\rangle
=0$, whereas the out-vacuum is defined by the outgoing states
$^{-}a_{l}\left| {\rm out},{\rm vac}\right\rangle =^{+}a_{k}\left|
{\rm out},{\rm vac}\right\rangle =0$, where $a$ are the particle
creation and annihilation operators, respectively \citep{So82}.
The number of the outgoing particles in the mode $k$ created in
the in-vacuum is $N_{kk^{\prime }}\equiv \left\langle {\rm
vac},{\rm in}\right| ^{+}a_{k}^{+}$\ $^{+}a_{k^{\prime }}\left|
{\rm vac},{\rm in}\right\rangle =\sum_{l}\left| \beta _{kl}\right|
^{2}\delta \left( k-k^{\prime }\right) $, where the coefficients
$\beta _{kl}$ are the elements of the single-particle $S$ matrix.
Due to the validity of the energy and momentum conservation laws,
one can generally choose $\beta _{kl}=\beta _{k}\delta _{kl}$ and
thus the number of created particles in the mode $k$ is simply
given by $N_{kk^{\prime }}=\left| \beta _{k}\right| ^{2}\delta
\left( k-k^{\prime }\right) $. The delta function can be
re-expressed as a delta function over the frequencies of the
associated modes, $\beta _{kl}=\mathcal T_{k}\delta \left( \omega
_{l}-\omega _{k}\right) $. Then the rule $\left\{ \delta \left(
\omega _{l}-\omega _{k}\right) \right\} ^{2}\rightarrow \left(
1/2\pi \right) \delta \left( \omega _{l}-\omega _{k}\right) \int
dt$ may be used to calculate a continuous rate of creation of
particle-antiparticle pairs \citep{So82,Kim02},
\begin{equation}
w=\frac{d}{dt}\left\langle N\right\rangle =\frac{1}{2\pi }\int
\left| \mathcal T_{k}\right| ^{2}d\omega .
\end{equation}

The physical meaning of the particle creation process is the
following: a wave packet of negative frequencies incident on an
electric field oriented along the $z-$direction will be partly
reflected by the electric field and partly transmitted to
$z\rightarrow \infty $ as a wave with positive frequencies. This
process is nothing else than tunnelling \citep{So82,Gr85,Wa88}.
However, in the following we restrict the tunnelling probability
to the transmission probability through the potential barrier but
exclude any non-zero transmission probability above the potential
barrier.

In the WKB
approximation the transmission probability $\mathcal T$ can be approximately given as $%
\left| \mathcal T\right| ^{2}=\exp \left( -2\int_{barrier}\left|
k_{z}\right| dz\right) =\exp\left(-2\sigma \right)$, where $\sigma
= \int_{barrier}\left| k_{z}\right| dz$ and $k_z$ is the
longitudinal momentum of the electron \citep{So82,Wa88,Kim02}.

For a fixed frequency $\omega $, the momentum $k_{z}$ of an
electron in the electrosphere of the quark stars is given by
\begin{equation}
k_{z}=\sqrt{m_{e}^{2}+k_{\perp }^{2}-\left[ \omega -V\left(
z\right) \right] ^{2}},
\end{equation}
where $k_{\perp }^{2}=k_{x}^{2}+k_{y}^{2}$ \citep{So82,Kim02}. The
mean particle production number $\left\langle N\right\rangle$ is
given by summing over all modes \citep{Gr85},
\begin{equation}
\left\langle N\right\rangle =\int e^{-2\sigma }\frac{dtd\omega }{2\pi }\frac{dxdk_{x}}{%
2\pi }\frac{dydk_{y}}{2\pi }.
\end{equation}

In order to estimate the electron-positron pair production rate in
the electrosphere, we have to find first the thickness of the
classically forbidden zone, or, equivalently, the limits of
integration $z_{\pm }$ in the transmission probability. They can
be obtained as solutions of the equation $k_{z}\left( z\right) =0$
and are given by
\begin{equation}
z_{\pm }=\frac{1}{2}\sqrt{\frac{3}{\alpha \pi }}\frac{1}{T}{\rm arcsinh}\left(\frac{%
\sqrt{2}\pi T}{\omega \mp \sqrt{m_{e}^{2}+k_{\perp
}^{2}}}\right)-z_{0}.
\end{equation}

Therefore, by taking into account the form of the electrostatic
potential in the electrosphere we obtain for $\sigma $ the
expression
\begin{equation}
\sigma =\int_{z_{-}}^{z_{+}}\sqrt{m_{e}^{2}+k_{\perp }^{2}-\left\{ \omega -%
\frac{\sqrt{2}\pi T}{\sinh \left[ 2\sqrt{\frac{\alpha \pi
}{3}}T\left( z+z_{0}\right) \right] }\right\} ^{2}}dz.
\end{equation}

With the help of the transformation $\eta =\left(
1/\sqrt{m_{e}^{2}+k_{\perp
}^{2}}\right) \left\{ \omega -\sqrt{2}\pi T/\sinh \left[ 2\sqrt{\alpha \pi /3%
}T\left( z+z_{0}\right) \right] \right\} $ we obtain the following
integral representation for $\sigma $:
\begin{equation}\label{sigma}
\sigma =\sqrt{\frac{3\pi }{2\alpha }}\frac{m_{e}^{2}+k_{\perp
}^{2}}{\omega
^{2}}\int_{-1}^{1}\frac{\sqrt{1-\eta ^{2}}d\eta }{\left( \frac{\sqrt{%
m_{e}^{2}+k_{\perp }^{2}}}{\omega }\eta -1\right) \sqrt{\frac{2\pi ^{2}T^{2}%
}{\omega ^{2}}+\left( \frac{\sqrt{m_{e}^{2}+k_{\perp
}^{2}}}{\omega }\eta -1\right) ^{2}}}.
\end{equation}

The variation of the transmission probability $\left| \mathcal
T\right| ^{2}=\exp
\left( -2\sigma \right) $ is represented, as a function of the parameter $%
\zeta =\sqrt{m_{e}^{2}+k_{\perp }^{2}}/\omega $ and for different
values of the temperature and particle energy ratio $2\pi
^{2}T^{2}/\omega ^{2}$ in
Fig. 2. For large values of the particle energy, $\omega >>\sqrt{%
m_{e}^{2}+k_{\perp }^{2}}$, $\zeta \rightarrow 0$ and the
transmission probability is equal to $1$,  $\left| \mathcal
T\right| ^{2}\rightarrow 1$. The transmission probability also
increases with the temperature of the electrosphere.

In the process of particle production in the electrosphere of
quark stars due to the tunnelling from the negative energy state,
a positive energy particle (an electron) is created, leaving a
hole in the negative energy continuum, which can be taken to be an
anti-particle (a positron) moving in a direction opposite to the
direction of the created particle. The created pair of particles
is characterized by the energy (frequency) $\omega $. For such a
pair, one cannot, strictly speaking, specify a particular point as
the location where the pair is produced \citep{Wa88}. One can only
say that the pair of particles begins to emerge between the point
$z_{-}$ and $z_{+}$. Nevertheless, one can associate the point
$z$, which is the solution of the equation $\omega =V(z)$, as the
location in the vicinity of which a pair of particles is produced.
With this approximate association, the energy of the produced
particle is then identified by an approximate location, and the
energy interval also can be approximately related to a spatial
interval at which the pair of particles is produced via the
relation $d\omega =\left( \partial V/\partial z\right) dz=eE\left(
z,T\right) dz$.

Moreover, we shall introduce polar coordinates in the momentum
space so that $k_{x}=k\cos \theta $ and $k_{y}=k\sin \theta $.
Therefore the electron-positron pair production rate $\dot{n_{\pm
}}$, giving the number of electron-positron pairs created per unit
time and per unit volume by the
electric field of the electrosphere at the surface of quark stars, $%
\left\langle N\right\rangle /\Delta t\Delta x\Delta y\Delta z$,
can be written as
\begin{eqnarray}\label{prode}
\dot{n_{\pm }}\left( z,T\right)  &=&\frac{s}{4\pi ^{2}}eE\left(
z,T\right) \times \nonumber\\
&&\int_{0}^{\infty }k\exp \left[ -2\sqrt{\frac{3\pi }{2\alpha }}\frac{%
m_{e}^{2}+k^{2}}{V^{2}\left( z,T\right) }\int_{-1}^{1}\frac{\sqrt{1-\eta ^{2}%
}\left( \frac{\sqrt{m_{e}^{2}+k^{2}}}{V\left( z,T\right) }\eta
-1\right)
^{-1}d\eta }{\sqrt{\frac{2\pi ^{2}T^{2}}{V ^{2}\left( z,T\right)}+\left( \frac{%
\sqrt{m_{e}^{2}+k^{2}}}{V\left( z,T\right) }\eta -1\right) ^{2}}}%
\right] dk,
\end{eqnarray}
where $s$ describes the spin degrees of freedom of the produced
particles ($s=1$ for bosons and $s=2$ for fermions).

Eq. (\ref{prode}) has a general formal structure which is very
similar to that of Eq. (\ref{prod}), describing the
electron-positron pair production in a constant electric field. In
both cases the pair production rate is proportional to the
intensity of the electric field, multiplied by an integral over
the transverse momentum of the particles. However, since in the
electrosphere $E(z,T)\rightarrow 0$ for large $z$, the particle
production rate also naturally tends to zero and outside the
electrosphere or in the regions with low electric field the
particle production ceases. The variation of the pair production
rate in the electrosphere, as a function of the distance $z$ is
represented for different values of the temperature in Fig. 3.

By introducing a new variable $\zeta
=\sqrt{m_{e}^{2}+k^{2}}/V\left( z,T\right) $, the pair production
rate can also be represented as
\begin{eqnarray}
\dot{n}\left( z,T\right)  &=&\frac{s}{4\pi ^{2}}eE\left(
z,T\right) V^{2}\left( z,T\right) \times  \nonumber\\
&&\int_{\frac{m_{e}}{V(z,T)}}^{\infty }\zeta \exp \left[
-2\sqrt{\frac{3\pi }{2\alpha }}\zeta
^{2}\int_{-1}^{1}\frac{\sqrt{1-\eta ^{2}}\left( \zeta \eta
-1\right) ^{-1}d\eta }{\sqrt{\frac{2\pi ^{2}T^{2}}{V\left(
z,T\right) ^{2}}+\left( \zeta \eta -1\right) ^{2}}}\right] d\zeta
.
\end{eqnarray}

In order to find an approximate representation of the pair
production rate we power expand the integrand in Eq.
(\ref{sigma}). In the first order in $\eta $ we obtain
\begin{eqnarray}
\frac{\sqrt{1-\eta ^{2}}}{\left( \frac{\sqrt{m_{e}^{2}+k_{\perp }^{2}}}{%
\omega }\eta -1\right) \sqrt{\frac{2\pi ^{2}T^{2}}{\omega ^{2}}+\left( \frac{%
\sqrt{m_{e}^{2}+k_{\perp }^{2}}}{\omega }\eta -1\right) ^{2}}}
&\approx
&\frac{\omega }{\sqrt{2\pi ^{2}T^{2}+\omega ^{2}}}+\frac{\sqrt{%
m_{e}^{2}+k_{\perp }^{2}}}{\sqrt{2\pi ^{2}T^{2}+\omega
^{2}}}\times  \nonumber\\
&&\left( \frac{\omega ^{2}}{2\pi ^{2}T^{2}+\omega ^{2}}+1\right)
\eta +O\left( \eta ^{2}\right).
\end{eqnarray}

Therefore in the first order of approximation $\sigma $ is given
by
\begin{equation}
\sigma \approx \sqrt{\frac{3\pi }{\alpha }}\frac{%
m_{e}^{2}+k_{\perp }^{2}}{\omega \sqrt{\frac{\omega ^{2}}{2}+\pi
^{2}T^{2}}}.
\end{equation}

By fixing the electron energy so that $\omega =V(z,T)$ and taking
into account the expression for the electric field in the
electrosphere, given by Eq. (\ref{el}), it follows that the
electron-positron pair production rate can be written as
\begin{eqnarray}
\dot{n}_{\pm }\left( z,T\right)  &\approx &\frac{2}{4\pi
^{2}}eE\left(
z,T\right) \int_{0}^{\infty }\exp \left( -2\sqrt{\frac{3\pi }{\alpha }}\frac{%
m_{e}^{2}+k^{2}}{V\sqrt{\frac{V^{2}}{2}+\pi ^{2}T^{2}}}\right)
kdk  \nonumber\\
&&\approx\frac{2}{4\pi ^{2}}eE\left( z,T\right) \int_{0}^{\infty
}\exp
\left[ -\pi \frac{m_{e}^{2}+k^{2}}{eE(z,T)}\right] kdk  \nonumber\\
&\approx &\frac{1}{4\pi ^{3}}e^{2}E^{2}(z,T)\exp \left[ -\pi \frac{E_{cr}}{%
E(z,T)}\right] .
\end{eqnarray}

This is exactly the leading term in the Schwinger formula for the
pair creation in a constant electric field $E_0$, but with the
constant electric field substituted with the inhomogeneous,
$z$-dependent electric field in the electrosphere, $E_0\rightarrow
E(z,T)$. Therefore, in the first approximation, it is possible to
study electron-positron pair production in arbitrary electric
fields in the electrosphere by simply substituting the constant
electric field in the Schwinger formula with the inhomogeneous
electric field.

In the second order of approximation we obtain
\begin{eqnarray}
\sigma  &\approx &\sqrt{\frac{3\pi }{\alpha }}\frac{5}{6}\frac{%
m_{e}^{2}+k^{2}}{\omega \sqrt{\frac{\omega ^{2}}{2}+\pi ^{2}T^{2}}}+\frac{1}{%
3}\sqrt{\frac{3\pi }{\alpha }}\frac{\left( m_{e}^{2}+k^{2}\right) ^{2}}{%
\omega \sqrt{\frac{\omega ^{2}}{2}+\pi ^{2}T^{2}}}\times
\nonumber\\
&&\left[ \frac{1}{\omega ^{2}}+\frac{1}{4}\frac{1}{\frac{\omega
^{2}}{2}+\pi ^{2}T^{2}}+\frac{3}{8}\frac{\omega ^{2}}{\left(
\frac{\omega ^{2}}{2}+\pi ^{2}T^{2}\right) ^{2}}\right] ,
\end{eqnarray}
giving
\begin{equation}
\dot{n}_{\pm }\left( z,T\right) \approx \frac{3}{8\pi
^{2}}e^{2}E^{2}\left(
z,T\right) \exp \left[ \frac{25\pi }{48\lambda \left( z,T\right) }\right] %
\left[ 3\lambda \left( z,T\right) \right] ^{-1/2}\left[ 1-{\rm
erf} \left( \chi \left( z,T\right) \right) \right] ,
\end{equation}
where $\lambda (z,T)=eE\left[ 1/V^{2}+\left( 1/3\pi \right)
V^{2}/E^{2}+\left( 2/3\pi ^{2}\right) V^{6}/E^{4}\right] $, $\chi =\sqrt{%
\lambda (z,T)/3\pi }E_{cr}/E+5\sqrt{\pi }/4\sqrt{3}\lambda \left(
z,T\right) $ and ${\rm erf}\left( x\right) =\left( 2/\sqrt{\pi
}\right) \int_{0}^{x}\exp \left( -u^{2}\right) du$.

\section{Boundary effects in pair production in the electrosphere }

The electron-positron pair creation in the electrosphere takes
place very close to the surface of the quark star, which
represents the boundary of the system. Therefore electron-positron
pair creation is localized to the half-space $z\geq 0$, and the
pair creation takes place in electric field localized to a bounded
region in the space. Boundary effects in electron-positron pair
creation by electric fields confined in a finite region of space
have been previously investigated, with the general result that
finite-size effects induce large deviations of the production rate
from what one deduces from the Schwinger formula
\citep{Wa88,Ma88,Ma89}.

In analyzing the surface effects on pair production there are two
very different situations, which lead to very different results.
The surface effects essentially depend on the orientation of the
electric field with respect to the boundary \citep{Ma88,Ma89}. In
the case of an electric field parallel to the boundary, the pair
production rate per unit time and unit volume in the region $z\geq
0$ is $w\left( z\right) =\sum_{n=1}^{\infty }w_{n}\left[ 1-\exp
\left( -eEz^{2}/n\pi \right) \right] $, where $w_{n}=\left(
e^{2}E^{2}/4\pi ^{3}\right) \exp \left( -n\pi E_{cr}/E\right) $ is
the Schwinger production rate in constant field with infinite
extension. As one can see for $z=0$ the boundary effects cancels
the Schwinger volume term, and thus electron-positron pair
creation cannot take place on a boundary parallel to the field
\citep{Ma88,Ma89}. Generally, there is a significant reduction of
the Schwinger pair production rate near a surface which is
parallel to the electric field. For an electric field of the order
of $E=5\times E_{cr}$, the reduction in the production rate at a
distance of $z=197.5$ fm from the parallel boundary is, for the case of the dominant term with $n=1$, $%
\left[ 1-\exp \left( -eEz^{2}/\pi \right) \right] =0.33$, while for $%
z=987.362$ fm the reduction rate is $0.999$. For the case of an
electric field $E=50\times E_{cr}$, we have a reduction rate of
$0.98$ at $z=197.5$ fm and $1 $ at $z=987.362$ fm.

However, the case of the electric field parallel to the boundary
is not relevant for the quark star electrosphere, in which the
electric field, which is oriented outward, is perpendicular on the
surface of the star (the boundary). In the case of the electric
field perpendicular to the boundary one can find the correction
terms by using a method based on approximating the Green functions
in terms of classical paths. The first order correction to the
effective Lagrangian for particle production is \citep{Ma89}
\begin{equation}
L_{eff}^{(1)}\left( z\right) =-\frac{1}{8\pi }\int_{0}^{\infty }\frac{ds}{%
s^{2}}\left[ eE\coth \left( eEs\right) -\frac{1}{s}\right] \exp
\left[ -im_{e}^{2}s+ie\frac{E}{2}z^{2}\coth \left(
\frac{eEs}{2}\right) \right] .
\end{equation}

The first order correction $w_{1}(z)$ to the pair production rate
is given by $w_{1}\left( z\right) =2{\rm Im}L_{eff}^{(1)}(z)$, and
for $m_e\approx 0$  can be written in the form \citep{Ma89}
\begin{equation}
w_{1}(z)=-\frac{1}{4\pi ^{2}}e^{2}E^{2}\int_{1}^{\infty }\Phi
\left( s\right) \sin \left( \frac{eE}{2}z^{2}s\right) ds,
\end{equation}
where
\begin{equation}
\Phi \left( s\right) =\frac{1}{\left( s^{2}-1\right) \left( \ln \frac{s+1}{%
s-1}\right) ^{2}}\left( s+\frac{1}{s}-\frac{2}{\ln
\frac{s+1}{s-1}}\right) .
\end{equation}

The ratio of the total production rate $w=w_{0}+w_{1}$ and of the
Schwinger production rate $w_{0}=\left( e^{2}E^{2}/4\pi
^{3}\right) \exp \left( -\pi E_{cr}/E\right) $ is represented, for
different values of the electric field, in Fig. 4.

For large $z$ the function ${\rm Im}L_{eff}^{(1)}(z)$ has the
asymptotic behavior
\begin{equation}
{\rm Im}L_{eff}^{(1)}(z)\approx -\frac{1}{8\pi
^{2}}e^{2}E^{2}\frac{\sin
\left( \frac{eE}{2}z^{2}\right) }{\ln \left( \frac{eE}{2}z^{2}\right) }+O%
\left[ \frac{1}{\ln \left( \frac{eE}{2}z^{2}\right) }\right] ,
\end{equation}
while for small $z$ it is given by
\begin{equation}
{\rm Im}L_{eff}^{(1)}(z)\approx -\frac{e^{2}E^{2}}{48\pi }+\frac{eE^{2}}{%
8\pi ^{2}}\left[ \frac{eE}{2}z^{2}\right] c,
\end{equation}
where $c=1/3-\int_{1}^{\infty }\left[ s\Phi \left( s\right)
-1/3\right] ds\approx -0.1$ \citep{Ma89}.

From these equations and from Fig. 4 it follows that when the
boundary is perpendicular to the field the pair production rate
exhibits oscillations as a function of the distance to the
boundary. No such oscillations occur when the electric field is
parallel to the boundary. For the discussion of the physical
origin of this phenomenon and its possible implications see
\citet{Ma89}.

As we have shown in the previous Section, in the first
approximation we can substitute the inhomogeneous electric field
in the results derived for the constant field case. We shall
follow this approach in the study of the effect of the boundary
(the quark star surface) on the pair production in the
electrosphere. Hence we assume that the boundary effect in the
pair production rate is given by
\begin{equation}
w_{1}(z,T)\approx -\frac{1}{4\pi ^{2}}e^{2}E^{2}\left( z,T\right)
\int_{1}^{\infty }\Phi \left( s\right) \sin \left[ \frac{eE(z,T)}{2}z^{2}s%
\right] ds,
\end{equation}
which is a straightforward generalization of the constant field
case.

Therefore the total rate production $\dot{n}_{\pm }^{(b)}\left( z,T\right) =%
\dot{n}_{\pm }\left( z,T\right) +w_{1}(z,T)$ of the
electron-positron pairs in the electrosphere, with the boundary
contribution included, is given by
\begin{eqnarray}
\dot{n}_{\pm }^{(b)}\left( z,T\right)  &\approx &\frac{s}{4\pi
^{2}}eE\left( z,T\right) V^{2}\left( z,T\right) \times  \nonumber\\
&&\int_{\frac{m_{e}}{V(z,T)}}^{\infty }\sigma \exp \left[
-2\sqrt{\frac{3\pi }{2\alpha }}\sigma
^{2}\int_{-1}^{1}\frac{\sqrt{1-\eta ^{2}}\left( \sigma \eta
-1\right) ^{-1}d\eta }{\sqrt{\frac{2\pi ^{2}T^{2}}{V\left(
z,T\right)
^{2}}+\left( \sigma \eta -1\right) ^{2}}}\right] d\sigma -  \nonumber \\
&&\frac{1}{4\pi ^{2}}e^{2}E^{2}\left( z,T\right) \int_{1}^{\infty
}\Phi \left( s\right) \sin \left[ \frac{eE(z,T)}{2}z^{2}s\right]
ds.
\end{eqnarray}

The ratio of the total pair production rate in the electrosphere,
with the boundary effects included, and the Schwinger pair
production rate in a constant electric field is represented, for
different values of the temperature, in Fig. 5. As expected, the
boundary effects significantly reduce the creation of
electron-positron pairs.

\section{Electron-positron pair flux of the electrosphere}

For an electron in the electrosphere sitting near the Fermi
surface the energy  is given locally by $\epsilon =\mu
_{e}(z,T)-V(z,T)$. Since from
the boundary conditions it follows that $\mu _{e}(z,T)\rightarrow 0$ and $%
V(z,T)\rightarrow 0$ for $z\rightarrow \infty $, the energy of the
electrons satisfy the condition $\epsilon =0$. From this result it
immediately follows that $\partial \epsilon /\partial z=0$. Electron and positron pairs are created with an energy $%
\varepsilon _{\pm }=m_{e}+T$, which is greater than the energy of
the electrons in the electrosphere,  $\varepsilon _{\pm }>\epsilon
$.

We define the electron-positron pair emissivity $Q_{\pm }$ of the
electrosphere as the energy created per unit time and per unit
volume by the electric field, given by the product of the number
of pairs multiplied by the energy of each pair. The mathematical
definition of the electron-positron pair emissivity in vacuum is
\begin{equation}
Q_{\pm }(z,T)=\varepsilon _{\pm }\frac{\left\langle N\right\rangle
}{\Delta t\Delta x\Delta y\Delta z}=\varepsilon _{\pm }n_{\pm
}\left( z,T\right) =\left( m_{e}+T\right) n_{\pm }\left(
z,T\right) .
\end{equation}

$Q_{\pm }(z,T)$ is general a local quantity, depending on the
distance to the quark star surface and on the temperature in the
electrosphere. The variation of the electron-positron emissivity
is represented, for different values of the temperature and for a
fixed quark star surface potential, in Fig. 6. In the case of a
constant electric field the electron-positron emissivity is
$Q_{\pm }^{(0)}\left( T\right) =\varepsilon _{\pm }w_{0}=\left(
m_{e}+T\right) \left( e^{2}/4\pi ^{3}\right) E^{2}\exp \left(
-\pi E_{cr}/E\right) $. For an electric field $E=50\times E_{cr}$ we obtain $%
Q_{\pm }^{(0)}\left( T\right) =1.28\times \left( m_{e}+T\right) $,
which for $T=3.5$ MeV gives $Q_{\pm }^{(0)}\left( T\right) =5.12$
MeV$^{5}$.

The electron-positron pair flux of the electrosphere in vacuum is
defined as
\begin{equation}
F_{\pm }^0\left( T\right) =\frac{1}{\pi }\int_{0}^{\infty }Q_{\pm }(z,T)dz=%
\frac{1}{\pi }\int_{0}^{\infty }\varepsilon _{\pm }n_{\pm }\left(
z,T\right) dz=\frac{\left( m_{e}+T\right) }{\pi }\int_{0}^{\infty
}n_{\pm }\left( z,T\right) dz.
\end{equation}

In the case of a constant electric field $E=50\times E_{cr}$, and
assuming that the electroshere extends to up to $d=500$ fm, we
obtain $F_{\pm }^{0}\left( T\right) =\left( m_{e}+T\right) \left(
e^{2}/4\pi ^{4}\right) E^{2}\exp \left( -\pi E_{cr}/E\right)
d\approx 1\times \left( m_{e}+T\right)
$, which for $T=3.5$ MeV gives $F_{\pm }^{0}\left( T\right) \approx 4$ MeV$%
^{4}$. The variation of the electron-positron flux is represented,
as a function of the temperature and for different values of the
quark star surface potential, in Fig. 7.

The variation of the electron-positron emissivity in the presence
of a boundary perpendicular to the electric field is represented,
as a function of the distance $z$ to the star's surface, in Fig.
8. The boundary of the star strongly suppresses the pair
production and induces an important qualitative change in the
behavior of the emissivity $Q$. The electron-positron flux in the
electrosphere, by taking into account the boundary effects, is
represented in Fig. 9. There is a significant effect in the
electron-positron flux due to the suppression of pair creation by
the quark star's surface.

The possibility of the thermal contribution to the pair creation
process was discussed by \citet{Gi99}. In the first loop
approximation there is no thermal contribution to the imaginary
part of the effective Lagrangian and therefore to the particle
production rate. This means that the pair production rate in
strong electric fields is basically independent of temperature,
and the temperature dependence of this effect in the electrosphere
of the quark stars is indirect and is due to the temperature
dependence of the electron number density and of the corresponding
electric field.

Finally, we compare the generalized Schwinger pair production
rates and the corresponding flux with the expressions proposed by
\citet{Us98a,Us98b}, and given by Eq. (\ref{usov}). We adopt a
typical model for the electrosphere with an
electron number density $n_{e}$ of the order $n_{e}=2\times 10^{-5}$ fm$%
^{-3}\approx 154$ MeV$^{3}$. The chemical potential of the
electrons, $\mu _{e}$, which is also equal to the Fermi energy of
the electrons $\epsilon _{F}$, is given by $\mu _{e}=\left( 3\pi
^{2}n_{e}\right) ^{1/3}\approx 16.58 $ MeV. The number density of
the pairs created by the electric field, which in this model is
taken to be equal to the density of electronic empty states with
energies below the pair creation threshold at thermodynamical
equilibrium, is given by $\Delta n_{\pm }^{(\rm Usov)}=\left(
3T/\mu _{e}\right) \exp \left( -2m_{e}/T\right) n_{e}\approx
27.8T\exp \left( -2m_{e}/T\right) $. By assuming that the electron
spectrum is thermalized due to electron-electron collisions, the
electron-positron pair flux from the electrosphere can be written
as $F_{\pm }^{({\rm Usov})}=0.01673\times T^{3}\times \left(
m_{e}+T\right) \exp \left( -2m_{e}/T\right) \times J\left( \xi
\right) $, where $\xi =1.59/T$. For $T=0.1$ MeV, $F_{\pm }^{({\rm
Usov})}\approx 7.4\times 10^{-9}$ MeV$^{4}$, $F_{\pm }^{({\rm
Usov})}\approx 0.53$ MeV$^{4}$ for $T=10$ MeV and $F_{\pm }^{({\rm
Usov})}\approx 3.45$ MeV$^{4}$ for $T=40$ MeV. As one can see from
Figs. 7 and 9, there is a very large difference between the
thermalized electron flux and the ''pure'' Schwinger flux due to
pair creation by the electric field.

However, this comparison is not correct in the concrete physical
framework of the quark star surface, since in the
\citet{Us98a,Us98b,Us01} model the electron thermalization time is
used to calculate the flux, while the Schwinger flux
is estimated in terms of the pair creation time in vacuum, which is of the order of $%
\Delta t\approx 1/\left( eE\right) ^{1/2}$ \citep{Ni70}. In order
to compare our results with the previous estimations of the
electron-positron pair rate at the surface of strange stars, we
have to systematically take into account the number of available
(free) electron states at the surface of the star and the
thermalization effects in both models.

 To compare the two mechanisms we start with the discussion
of the created particle number densities in the two models. In the
Schwinger mechanism in vacuum, the particle
number density is $\Delta n_{\pm }^0=\dot{n}\left( z,T\right) \Delta t=\dot{n}%
\left( z,T\right) /\left( eE\right) ^{1/2}$, which, in the
approximation of the constant electric field, can be written as
\begin{equation}\label{dens}
\Delta n_{\pm }^0=\frac{1}{4\pi ^{3}}m_{e}^{3}\left(
\frac{E}{E_{cr}}\right) ^{3/2}\exp \left( -\pi
\frac{E_{cr}}{E}\right) .
\end{equation}

For a fixed value of the electric potential in the electrosphere
$V_{q}$, which for simplicity we take as the electric potential at
the surface of the
quark star, the electric field is given by $E=\sqrt{4/3\pi }V_{q}\sqrt{%
V_{q}^{2}/2+\pi ^{2}T^{2}}$, and the electron number density is $%
n_{e}=V_{q}^{3}/3\pi ^{2}+V_{q}T^{2}/3$. The temperature dependence of $%
\Delta n_{\pm }$ is only through the temperature dependence of the
electric
field $E$, which is significant only for values of the temperature so that $%
T\geq V_{q}$.

However, the number of free electron states at the quark star
surface is given by $\Delta n_{\pm }^{({\rm Usov})}=3T\exp \left(
-2m_{e}/T\right) \left( V_{q}^{2}/3\pi ^{2}+T^{2}/3\right) $. From
a physical point of view, $\Delta n_{\pm }^{({\rm Usov})}$
represents the number of available quantum states for pair
creation \citep{Us98a,Us98b,Us01}. Moreover, these states are not
necessarily filled, because of the finite value of the electric
field. Hence, the number of created electron-positron pairs is
also finite, and this number can be less than the number of
available quantum states $\Delta n_{\pm }^{({\rm Usov})}$.
Therefore, electron-positron pair creation is possible once the
condition
\begin{equation}
\Delta n_{\pm }^{({\rm Usov})}\geq \Delta n_{\pm }^0,
\end{equation}
is fulfilled. Generally, these condition will be satisfied for
temperatures so that $T\geq T_{\rm cr}$, where the critical
temperature can be obtained from the condition
\begin{equation}
3T_{cr}\exp \left( -\frac{2m_{e}}{T_{cr}}\right) \left(\frac{
V_{q}^{2}}{3\pi ^{2}}+\frac{T_{cr}^{2}}{3}\right)=\frac{1}{4\pi
^{3}}m_{e}^{3}\left[ \frac{E\left(T_{cr}\right)}{E_{cr}}\right]
^{3/2}\exp \left[ -\pi \frac{E_{cr}}{E\left(T_{cr}\right)}\right]
\end{equation}

The value of $T_{\rm cr}$ depends on the electrostatic properties
of the quark star surface. The critical temperature is
represented, as a function of the electrostatic potential $V_q$ at
the quark star surface in Fig. 10. The critical temperature is
increasing with increasing $V_q$. Hence, when $T\leq T_{cr}$, all
quantum states are filled, whereas for $T> T_{cr}$ the number of
available quantum states may exceed the number of
electron-positron pairs created by the electric field.

Therefore, by taking into account the number of free electron
states we define the electron-positron pair number density $\Delta
n_{\pm }$ as
\begin{equation}
\Delta n_{\pm }=\left\{
\begin{array}{c}
\Delta n_{\pm }^{({\rm Usov})},T<T_{cr},\\
\Delta n_{\pm }^{0},T\geq T_{cr}.
\end{array}
\right.
\end{equation}

The variation of the particle number densities in the two models
is presented in Fig. 11. For the chosen values of $V_{q}$ the
condition $\Delta n_{\pm }^{({\rm Usov})}\geq \Delta n_{\pm }^0$
holds for temperatures $T\gtrapprox 0.1$ MeV for $V_q=5$ MeV and
$T\gtrapprox 0.4$ MeV for $V_q=15$ MeV, respectively. It should be
noted that despite the number of available free electron states
increases very rapidly with the temperature, due to the
temperature independence of the Schwinger process, the number of
the created electron-positron pairs is basically determined by a
single parameter, the quark star surface potential.

In the model of \citet{Us98a,Us01} the thermalized
electron-positron pair flux is given by
\begin{equation}
F_{\pm }^{{\rm (th)}}=4\pi R^{2}\Delta r_{E}\Delta n_{\pm }^{{\rm
(Usov)}}t_{th}^{-1},
\end{equation}
where $R$ is the radius of the star, $\Delta r_{E}$ the thickness
of the emitting region and $t_{th}^{-1}$
 is the characteristic time of thermalization of the electrons. For $%
t_{th}^{-1}$ the expression $t_{th}^{-1}\approx \left( 3/2\pi
\right) \left( \alpha /\sqrt{\pi }\right) \left( T^{2}/\varepsilon
_{F}\right) J\left( \xi \right) $, with $\xi =2\sqrt{\alpha /\pi
}\left( \varepsilon _{F}/T\right) $, was assumed
\citep{Us98a,Us01}. The function $J\left( \xi \right) $ is defined
by Eq. (\ref{j}).

In order to obtain a more realistic description of the thermalized
electron-positron flux of the electrosphere which takes into
account the inhomogeneities of the electron and electric field
distributions, we assume that, due to its dependence on the Fermi
energy $\varepsilon _{F}\approx \mu _e=V(z,T)$, the thermalization
time is also a function of the distance $z$ to the quark star
surface.

Therefore we define the thermalized electron-positron flux from
the quark star's surface in the generalized emission model as
\begin{equation}
F_{\pm }^{{\rm (th)}}=4\pi R^{2}\int_{0}^{\infty }\Delta n_{\pm
}\left( z,T\right)
t_{th}^{-1}\left( z,T\right) dz\approx 4\pi R^{2}\frac{3\alpha }{2\pi ^{3/2}}%
T^{2}\int_{0}^{\infty }\Delta n_{\pm }\left( z,T\right) \frac{J\left[ 2\sqrt{\frac{%
\alpha }{\pi }}\frac{V\left( z,T\right) }{T}\right] }{V\left(
z,T\right) }dz.
\end{equation}

The variations of the thermalized electron-positron fluxes in the
Usov model and in the generalized electron-positron emission
mechanism are presented in Fig. 12.

In the limit of low temperatures the condition $T/V_q\rightarrow
0$ holds with a very good approximation. By assuming that the
number density of the $e^{+}-e^{-}$ pairs can be approximated by
Eq. (\ref{dens}), in which all the parameters are estimated near
the quark star surface $z\approx 0$, the electron-positron
thermalized flux from the electrosphere of the quark stars can be
represented in an approximate form as
\begin{eqnarray}
F_{\pm }^{\mathrm{(th)}} &\approx &4\pi R^{2}\Delta r_{E}\times
V_{q}\left\{
\frac{\alpha \exp \left( -\sqrt{\frac{3}{2}}\pi ^{3/2}\frac{E_{cr}}{V_{q}^{2}%
}\right) m_{e}^{3}\left( \frac{V_{q}^{2}}{E_{cr}}\right) ^{3/2}}{%
83^{3/4}\left( 2\pi \right) ^{1/4}}\left( \frac{T}{V_{q}}\right)
^{2}\right. -   \nonumber\\
&&\left.0.7165\sqrt{\alpha }\exp \left( -\sqrt{\frac{3}{2}}\pi ^{3/2}\frac{E_{cr}}{%
V_{q}^{2}}\right) m_{e}^{3}\left( \frac{V_{q}^{2}}{E_{cr}}\right)
^{3/2}\left( \frac{T}{V_{q}}\right) ^{3}+...\right\}.
\end{eqnarray}

For $T=0$ the electron-positron flux from the quark star surface
is zero. Generally, $F_{\pm }$ is a function of the electrostatic
potential at the quark star surface $V_q$ and of the temperature
$T$.

The presence of a surface magnetic field $H$ can also enhance the
pair production rate \citep{Ni70}. The magnetic field increases
the pair production rate by a factor $\delta _H=\pi H/E\coth
\left( \pi H/E\right) $. If $H>>E$, there will be a significant
increase in the pair production rate. The electric field of the
electrosphere could be as high as $E=40E_{crit}\approx 120$
MeV$^2$. On the other hand the estimated magnetic fields at the
surface of the quark stars could be of the order of $H\approx
10^{15}-10^{17}$G $\approx 20-2000$ MeV$^2$ (1G $=1.953\times
10^{-14}$ MeV$^2$).

Magnetic fields with such high values may be present in very young
quark stars. Assuming equipartition of energy, the energy of the
differential rotation can be converted into magnetic energy, so
that $I\Omega ^{2}\left( \Delta \Omega /\Omega \right) \approx
\left( 4\pi /3\right) R^{3}\left( H^{2}/8\pi \right) $, where $I$
is the moment of inertia of the star, $R$ its radius and $\Omega $
and $\Delta \Omega $ are the angular velocity and the variation of
the angular velocity, respectively. Therefore the magnetic field
of a young quark star can be approximated as $H\approx
10^{4}\left( \Delta \Omega /\Omega \right) ^{1/2}$ MeV$^{2}$. By
assuming that $\Delta \Omega /\Omega \approx 0.03$, we can obtain
values of the magnetic field as high as $H\approx 2000$ MeV$^{2}$.
Of course magnetic fields of such strength are not stable, because
they will be pushed to and through the surface by buoyant forces
and then reconnect \citep{Klu98}. For a magnetic field of the
order of $H\approx 2000$ MeV$^{2}$ we have $\delta _H\approx 53$.

Therefore strong magnetic fields can significantly increase the
electron-positron pair production rate, and, consequently the
luminosity of the electrosphere of quark stars.

\section{Conclusions}

In the present paper we have re-considered the electron-positron
pair emission from the electrosphere of quark stars, as originally
proposed by \citet{Us98a,Us98b}, by pointing out the important
role the boundary effects and the inhomogeneity in the
distribution of the electric field may play in the pair creation
process. At zero temperature, there are no available free energy
states in the electron plasma at the strange star's surface.
Therefore, at low temperatures $T\leq T_{cr}\approx 0.1$ MeV
(corresponding to a quark star surface electric potential of
$V_q=5$ MeV), the pair production mechanism by the strong electric
field of the electrosphere is severely limited by the quantum
effects and the exclusion principle specific to the Fermi-Dirac
statistics. At high temperatures $T\geq T_{cr}\approx 0.1$ MeV,
the pair creation rate is controlled by the electric field $E$ and
not by the temperature, because such a process is essentially a
quantum process. Once the number of available electron states
becomes higher than the Schwinger pair production rate,
electron-positron pairs can be freely created by the electric
field at the surface of strange stars. This happens at a critical
temperature $T_{cr}$, which strongly depends on the electrostatic
properties of the quark star surface. The critical temperature
increases with the increase of the electrostatic potential $V_q$.
At high enough temperatures, the pair creation process is almost
independent of the temperature and is controlled exclusively by
the electric field. On the other hand, the actual thermalized pair
creation rate, which, from an astrophysical and observational
point of view is the most relevant quantity,  depends on the
temperature through the thermalized time scale.The number density
of pairs can be accurately evaluated by using the Schwinger
formalism and by taking into account the inhomogeneities in the
electric field distribution. However, the boundary effects also
induce large quantitative and qualitative deviations of the
particle production rate from what one deduces with the Schwinger
formula and its generalization for the inhomogeneous electric
field of the electrosphere.

Due to all these effects, we estimate that at high temperatures
the energy flux due to $e^{-}-e^{+}$ pairs production could be
lower than in the initial proposal of \citet{Us98a,Us98b}.
However, this flux could still be the main observational signature
of a quark star. On the other hand, the presence of a strong
magnetic field at the quark star surface may significantly enhance
the electron-positron flux.

The possible astrophysical and observational implications of the
direct pair production effect will be considered in a future
publication.

\section*{Acknowledgements}

The authors would like to thank the anonymous referee for very
helpful comments and suggestions. This work is supported by a RGC
grant of the government of the Hong Kong SAR.

\clearpage

\begin{figure}
\plotone{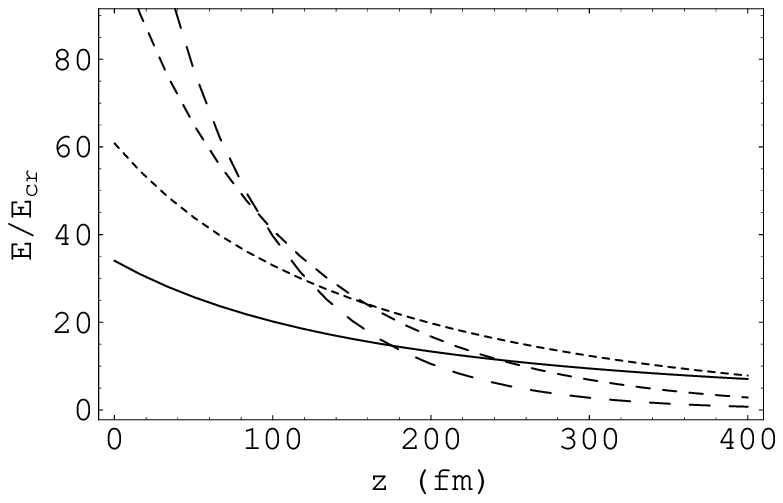} \caption{Ratio of the electric field of the
electrosphere $E$ and of the critical electric field
$E_{cr}=m_e^2/e$, as a function of the distance $z$ (fm), for
different values of the temperature: $T=0.01$ MeV (solid curve),
$T=5$ MeV (dotted curve), $T=10$ MeV (dashed curve) and $T=15$ MeV
(long dashed curve). In all cases $V_{q}(0,T)=15$ MeV.
\label{FIG1}}
\end{figure}

\begin{figure}
\plotone{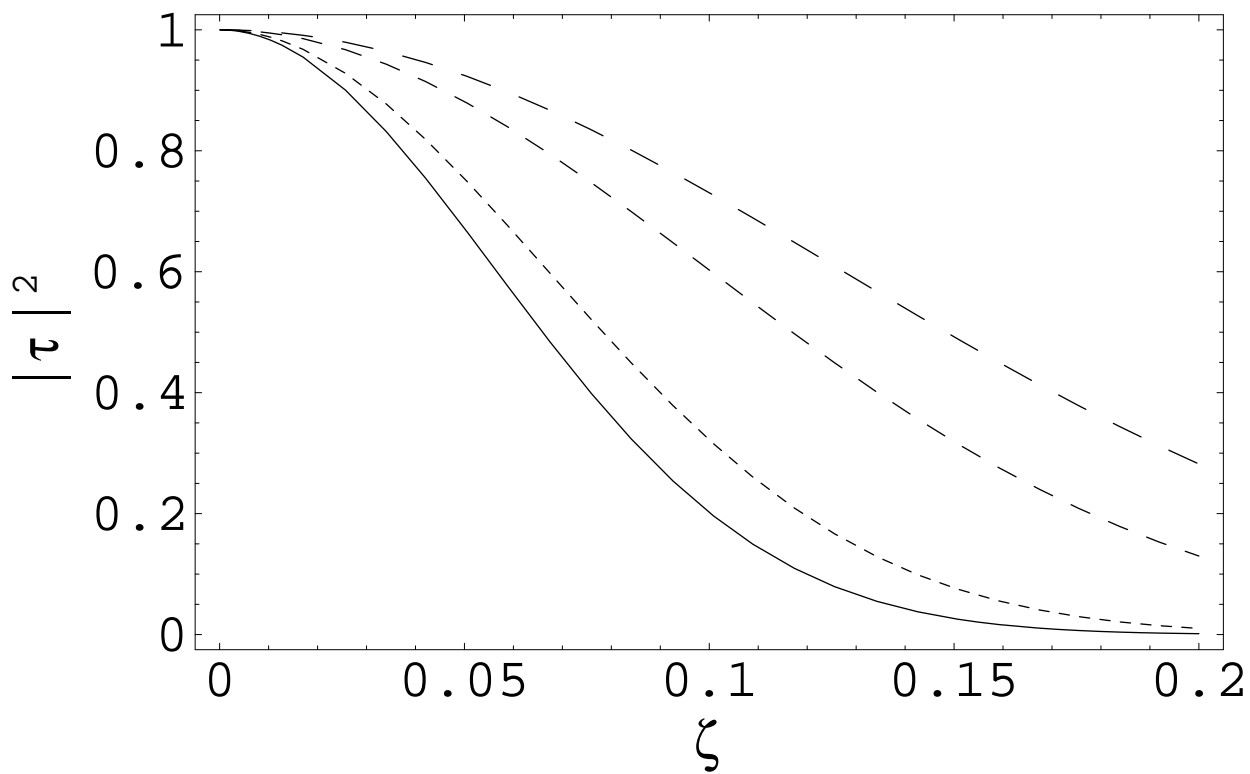} \caption{Transmission probability $\left|
\mathcal T\right| ^{2}=\exp \left( -2\sigma \right) $ for
electron-positron pair creation in the electrosphere of quark
stars
as a function of the parameter $%
\zeta =\sqrt{m_{e}^{2}+k_{\perp }^{2}}/\omega $ for different
values of the temperature energy ratio: $2\pi ^2T^2/\omega ^2=0.1$
(solid curve), $2\pi ^2T^2/\omega ^2=1$ (dotted curve), $2\pi
^2T^2/\omega ^2=3$ (dashed curve) and $2\pi ^2T^2/\omega ^2=5$
(long dashed curve). \label{FIG2}}
\end{figure}

\begin{figure}
\plotone{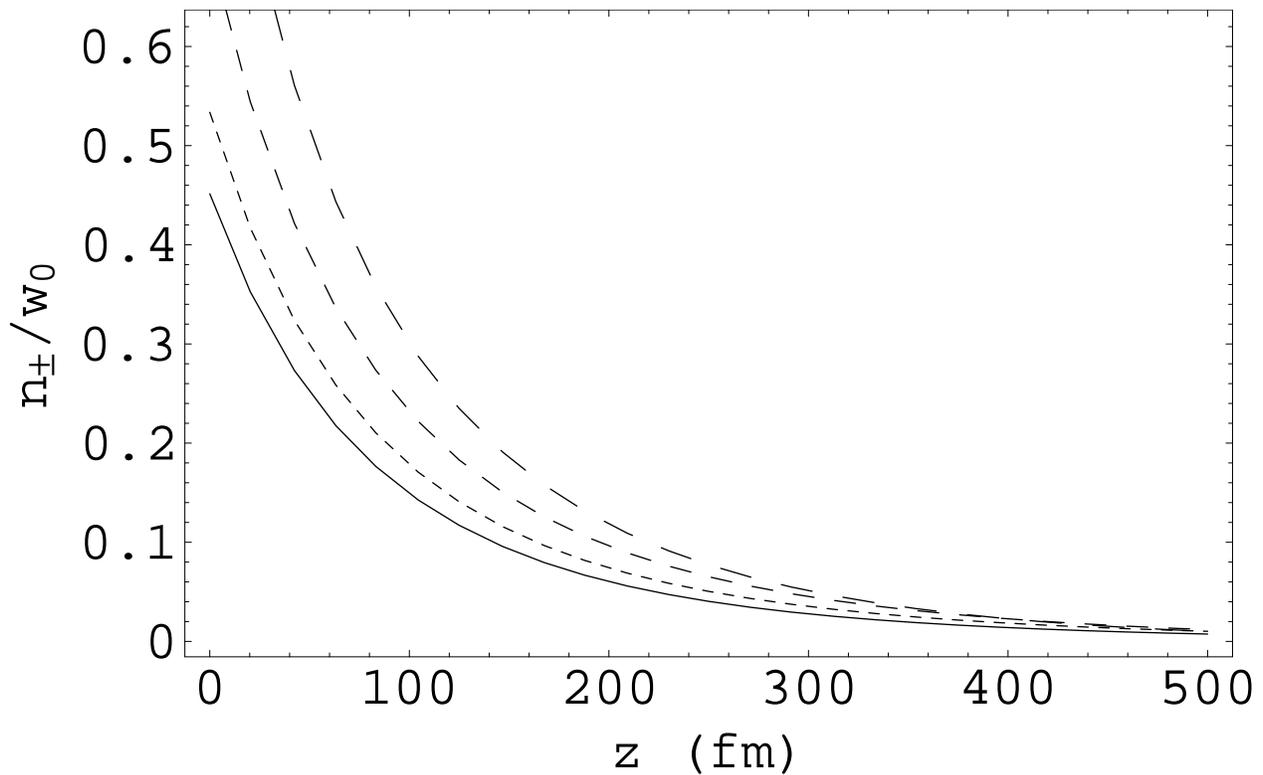} \caption{Ratio of the electron-positron pair
production rate $n_{\pm }$ in the electrosphere and of the
Schwinger rate $w_0$ for a constant electric field $E_0=50\times
E_{cr}$ as a function of the distance $z$ (fm) for different
values of the temperature: $T=0.5$ MeV (solid curve), $T=1.5$ MeV
(dotted curve), $T=2.5$ MeV (short dashed curve) and $T=3.5$ MeV
(long dashed curve). In all cases the surface electrostatic
potential of the quark star is $V_q=15$ MeV. \label{FIG3}}
\end{figure}

\begin{figure}
\plotone{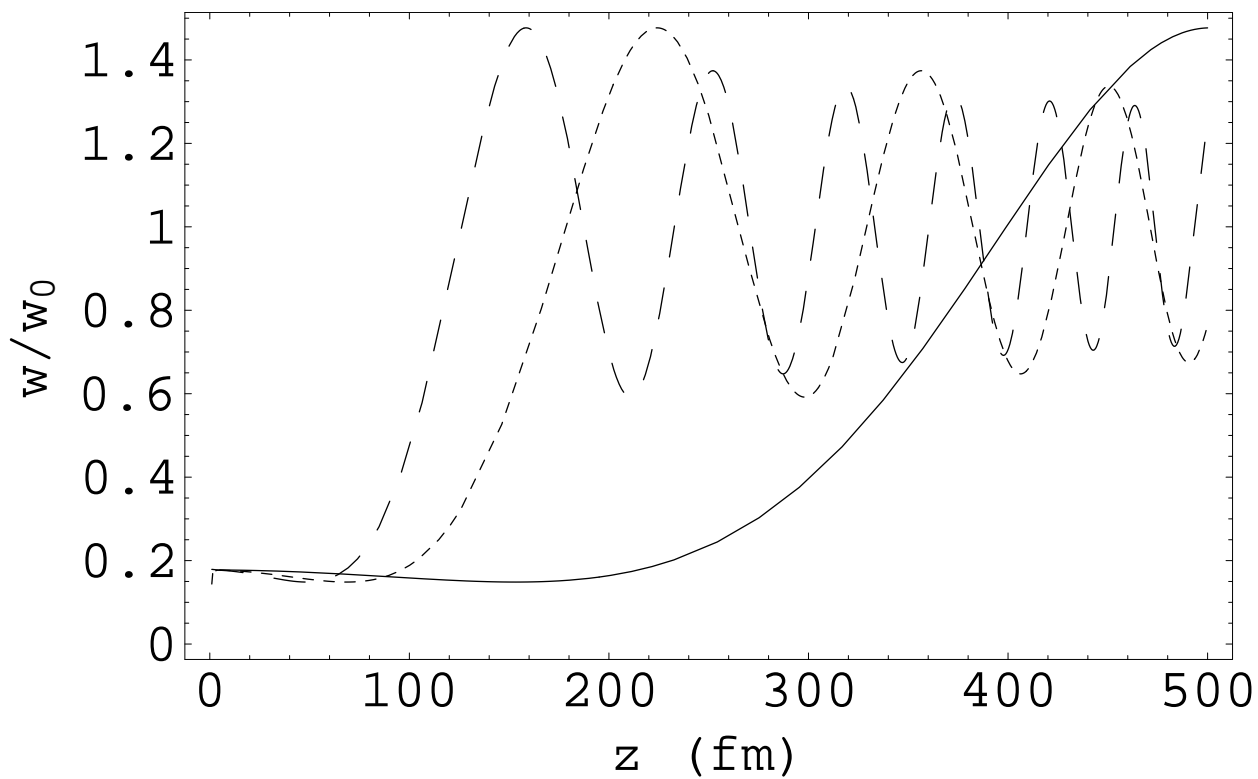} \caption{Ratio of the pair production rate
$w=w_0+w_1$ for a constant electric field, with the boundary
effects included, and the Schwinger rate $w_0$, as a function of
the distance $z$ (fm), for different values of the electric field:
$E=5\times E_{cr}$ (solid curve), $E=25\times E_{cr}$ (dotted
curve) and $E=50\times E_{cr}$ MeV (dashed curve). \label{FIG4}}
\end{figure}

\begin{figure}
\plotone{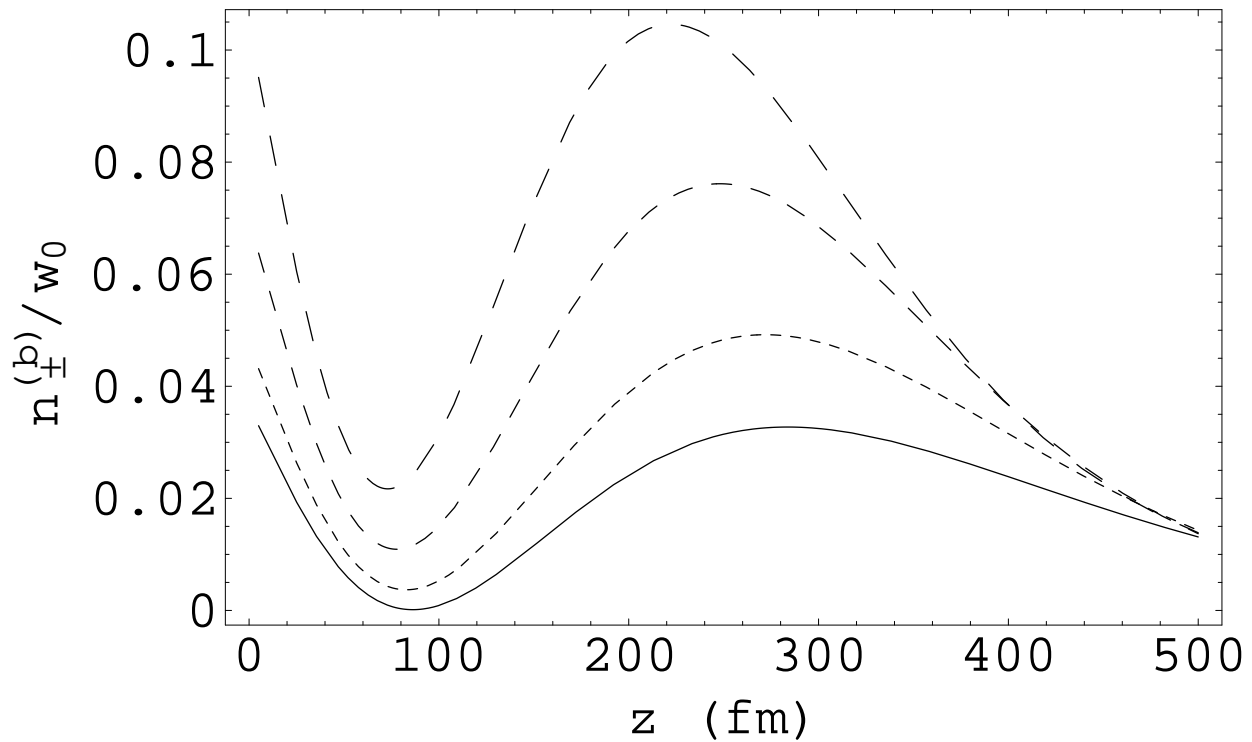} \caption{Ratio of the pair production rate
$n_{\pm }^{(b)}=n_{\pm }+w_1$ in the electrosphere of the quark
stars, with the boundary effects included, and the Schwinger rate
of pair creation in a constant electric field $w_0$, as a function
of the distance $z$ (fm), for different values of the temperature:
$T=0.5$ MeV (solid curve), $T=1.5$ MeV (dotted curve), $T=2.5$ MeV
( short dashed curve) and $T=3.5$ MeV (long dashed curve). The
value of the constant electric field used to calculate the
Schwinger rate is $E_0=50\times E_{cr}$. In all cases the surface
electrostatic potential of the quark star is $V_q=15$ MeV.
\label{FIG5}}
\end{figure}

\begin{figure}
\plotone{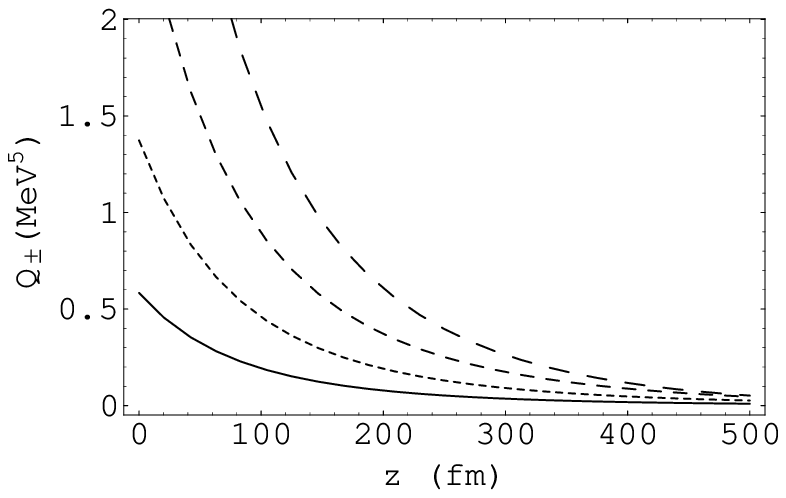} \caption{Electron-positron pair emissivity of the
electrosphere of the quark stars, as a function of the distance
$z$ (fm), for different values of the temperature: $T=0.5$ MeV
(solid curve), $T=1.5$ MeV (dotted curve), $T=2.5$ MeV ( short
dashed curve) and $T=3.5$ MeV (long dashed curve). In all cases
the surface electrostatic potential of the quark star is $V_q=15$
MeV. \label{FIG6}}
\end{figure}

\begin{figure}
\plotone{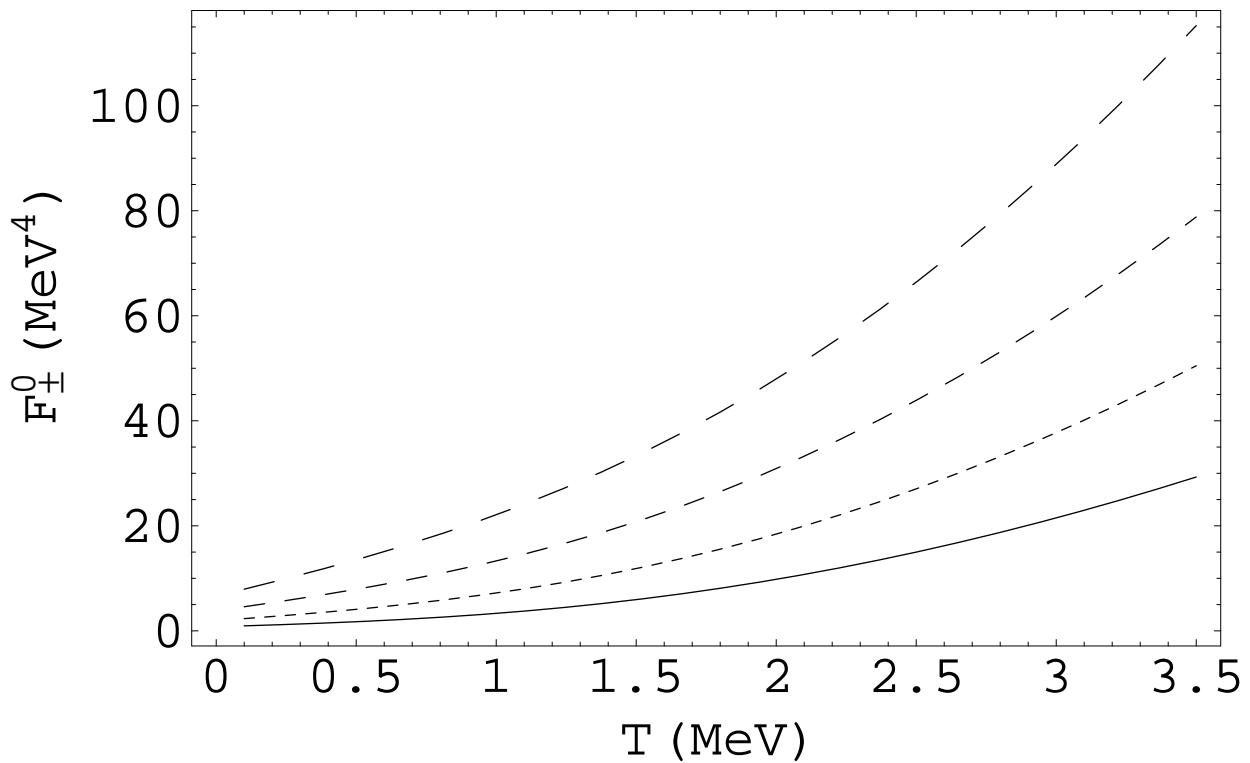} \caption{Vacuum electron-positron pair flux
$F_{\pm }^0$ of the electrosphere of the quark stars, as a
function of the temperature $T$ (MeV), for different values of the
surface electrostatic potential $V_q$: $V_q=8$ MeV (solid curve),
$V_q=10$ MeV (dotted curve), $V_q=12$ MeV ( short dashed curve)
and $V_q=14$ MeV (long dashed curve). \label{FIG7}}
\end{figure}

\begin{figure}
\plotone{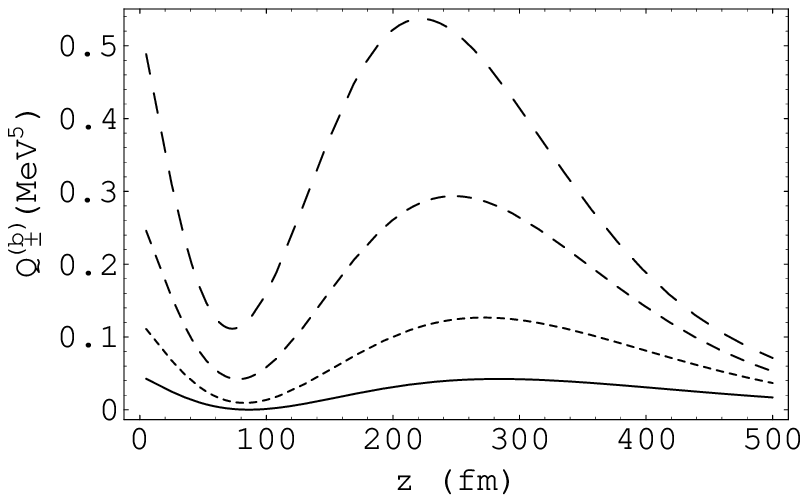} \caption{Vacuum electron-positron pair emissivity
$Q_{\pm }^{(b)}$ of the electrosphere of the quark stars, with the
boundary effects included, as a function of the distance $z$ (fm),
for different values of the temperature: $T=0.5$ MeV (solid
curve), $T=1.5$ MeV (dotted curve), $T=2.5$ MeV ( short dashed
curve) and $T=3.5$ MeV (long dashed curve). In all cases the
surface electrostatic potential of the quark star is $V_q=15$ MeV.
\label{FIG8}}
\end{figure}

\begin{figure}
\plotone{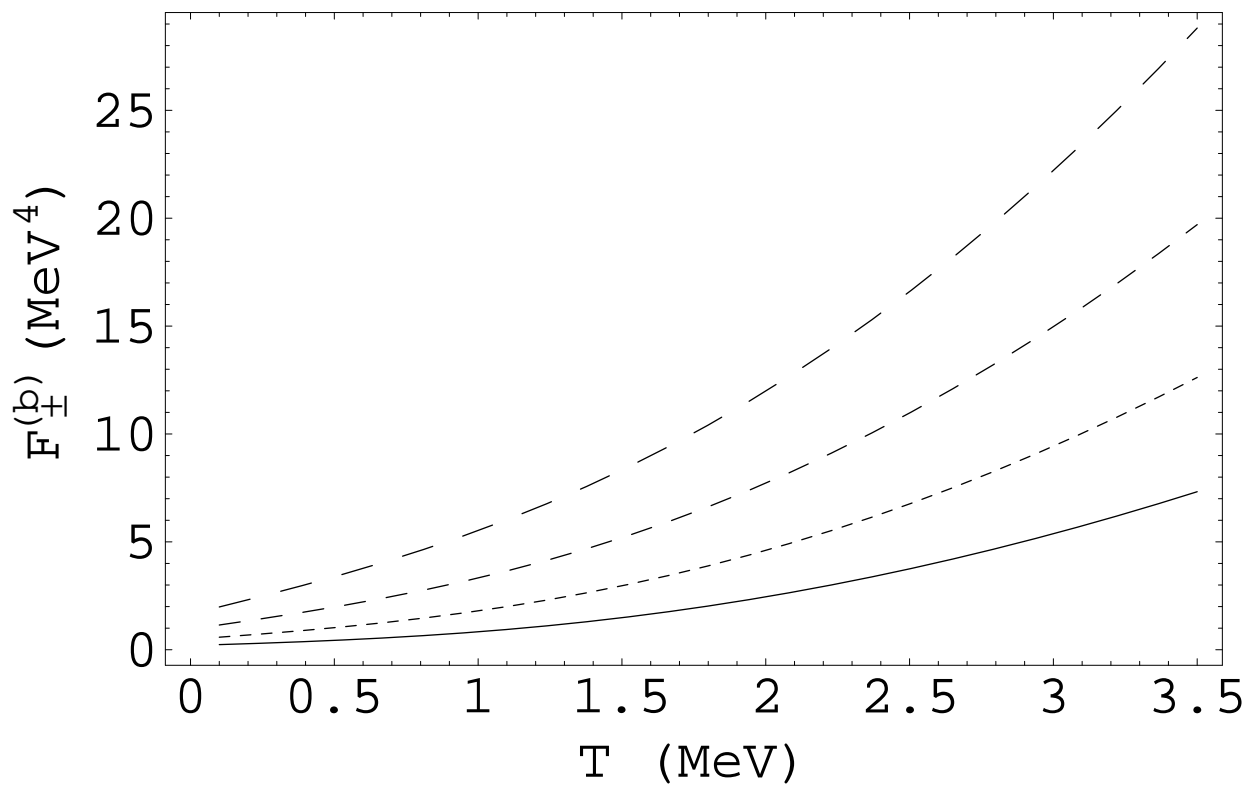} \caption{Vacuum electron-positron pair flux
$F_{\pm }^{(b)}$ of the electrosphere of the quark stars in the
presence of the boundary effects, as a function of the temperature
$T$ (MeV), for different values of the surface electrostatic
potential $V_q$: $V_q=8$ MeV (solid curve), $V_q=10$ MeV (dotted
curve), $V_q=12$ MeV ( short dashed curve) and $V_q=14$ MeV (long
dashed curve). \label{FIG9}}
\end{figure}

\begin{figure}
\plotone{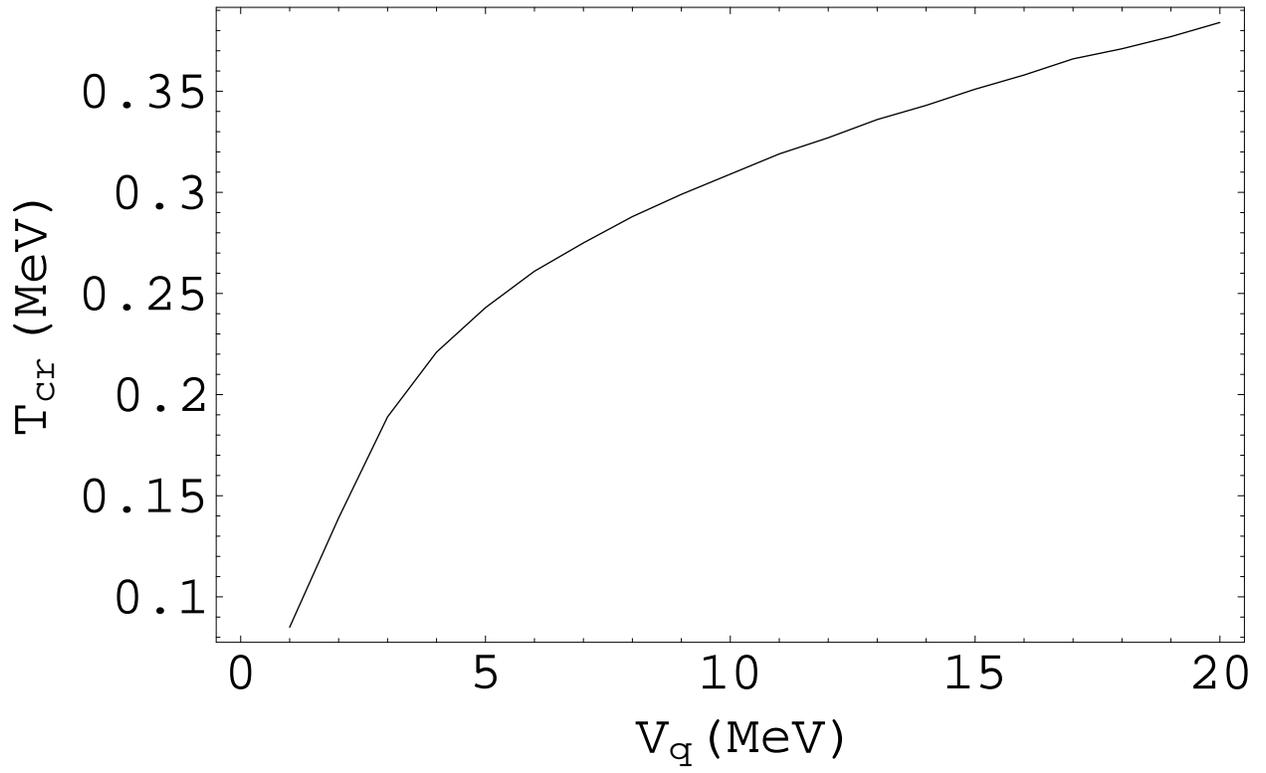} \caption{Critical temperature $T_{cr}$ as a
function of the surface electrostatic potential of the quark star.
\label{FIG10}}
\end{figure}

\begin{figure}
\plotone{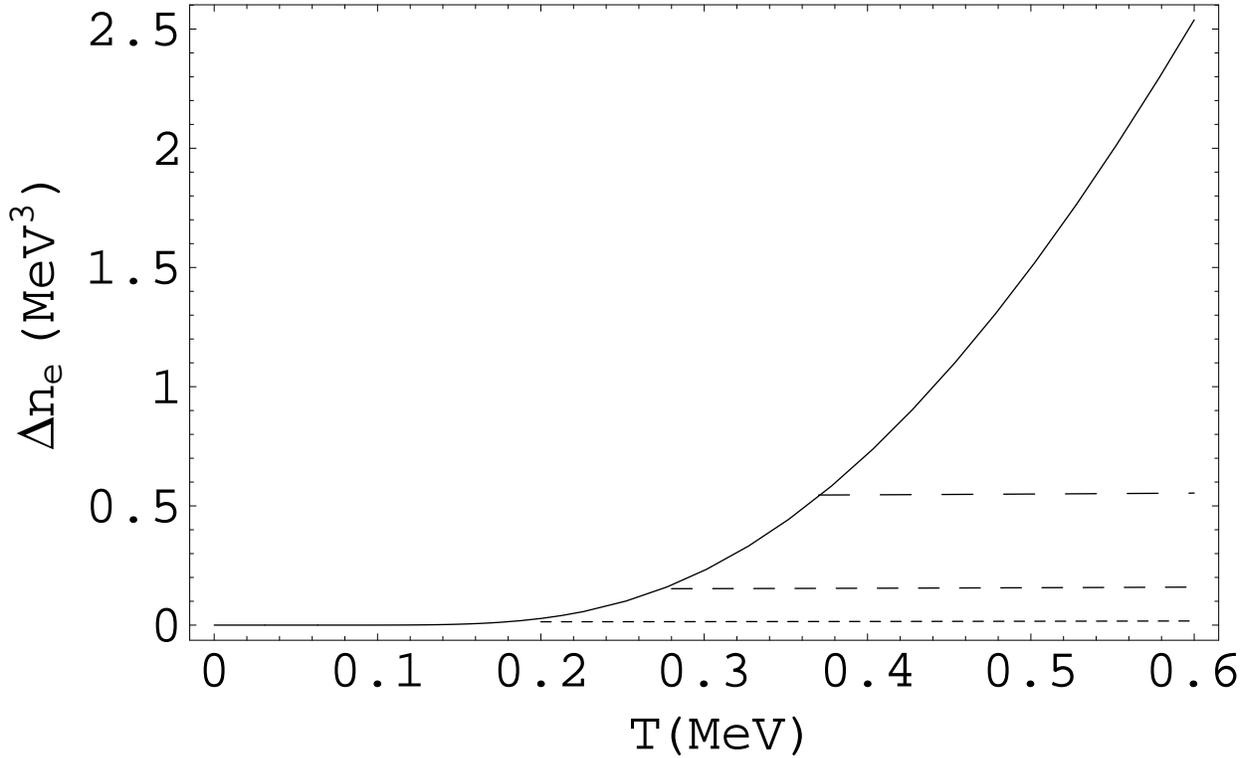} \caption{Temperature dependence of the
electron-positron number density $\Delta n_{\pm }^{({\rm Usov})}$
in the Usov mechanism (solid curve) and the particle number
density $\Delta n_{\pm }$ in the Schwinger mechanism  for
different values of the surface electrostatic potential: $V_q=5$
MeV (dotted curve), $V_q=10$ MeV (short dashed curve) and $V_q=15$
MeV (long dashed curve). For the Usov mechanism we have assumed
that the Fermi energy of the electrons is $\varepsilon _{F}=20$
MeV and the thickness of the electrosphere is $\Delta r_E=1000$
fm. \label{FIG10}}
\end{figure}

\begin{figure}
\plotone{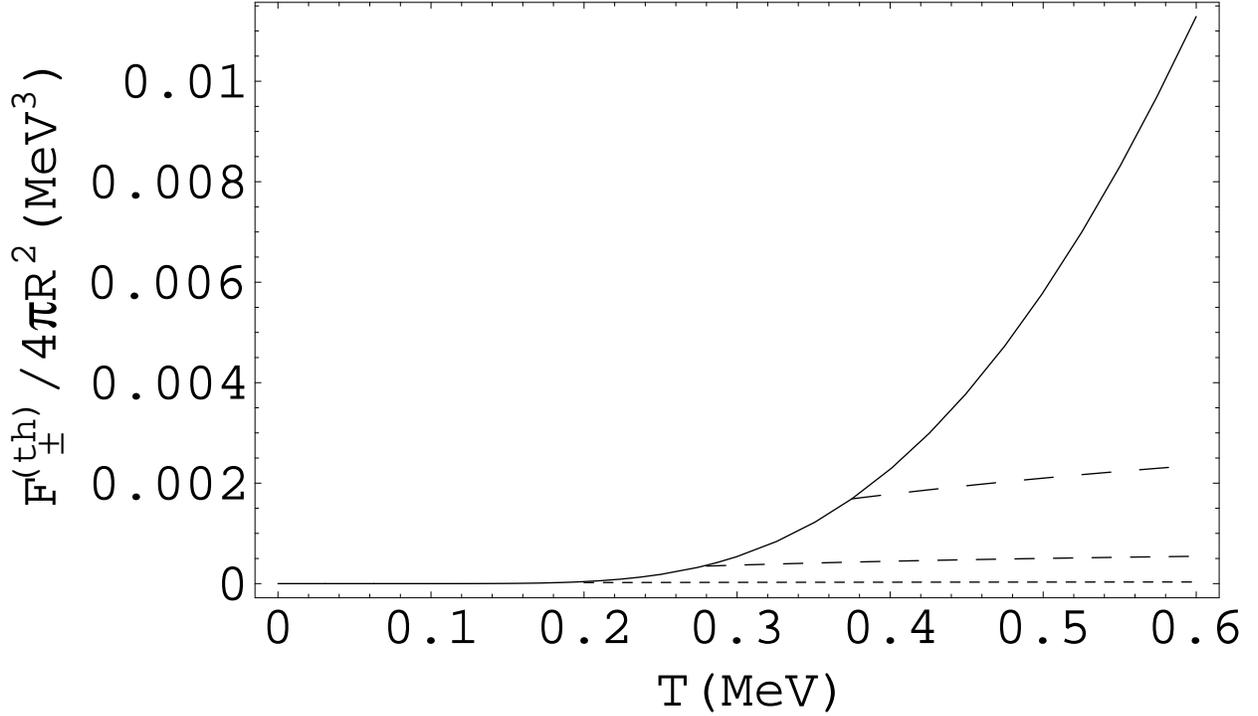} \caption{Temperature dependence of the
thermalized electron-positron flux  $F_{\pm }^{{\rm (th)}}/4\pi
R^{2}$ in the Usov mechanism  (solid curve) and in the generalized
electron-positron emission mechanism for different quark star
surface electrostatic potentials:  $V_q=5$ MeV (dotted curve),
$V_q=10$ MeV (short dashed curve) and $V_q=15$ MeV (long dashed
curve). For the Usov mechanism we have assumed that the Fermi
energy of the electrons is $\varepsilon _{F}=20$ MeV and the
thickness of the electrosphere is $\Delta r_E=1000$ fm.
\label{FIG11}}
\end{figure}


\end{document}